\documentclass[structabstract]{aa}
\usepackage{amssymb}
\usepackage{graphicx}
\usepackage{txfonts}
\begin{document}
   \title{The optical variability of flat-spectrum radio quasars in the SDSS Stripe 82 region}


   \author{Minfeng Gu
          \inst{1,2}
          \and
          Y. L. Ai\inst{3,4}
          }

   \institute{Key Laboratory for Research in Galaxies and Cosmology, Shanghai Astronomical Observatory,
    Chinese Academy of Sciences, 80 Nandan Road, Shanghai 200030, China\\
              \email{gumf@shao.ac.cn}
         \and
             Department of Physics, University of California, Santa Barbara, CA 93106, USA
         \and National Astronomical Observatories/Yunnan Observatory,
             Chinese Academy of Sciences, P.O. Box 110, 650011 Kunming, Yunnan, China
         \and
             Key Laboratory for the Structure and Evolution of Celestial Objects,
             Chinese Academy of Sciences, P.O. Box 110, 650011 Kunming, Yunnan, China
             }



\titlerunning{Optical variability of FSRQs in Stripe 82}
\authorrunning{M. F. Gu \& Y. L. Ai}

  \abstract
   {Although a bluer-when-brighter trend is commonly observed in blazars,
   the opposite trend of redder-when-brighter has also been found in
some blazars.}
   {We investigate the frequency of the redder-when-brighter trend in flat-spectrum radio quasars (FSRQs).}
   {We investigate the optical variability of 29 FSRQs in the SDSS
Stripe 82 region using SDSS DR7 released multi-epoch data covering
about nine years. We determined the spectral index by fitting a
powerlaw to SDSS ${ugriz}$ photometric data, and explored the
relationship between the spectral index and source brightness.}
   {For all FSRQs studied, we detect variations in $r$ band flux of overall amplitude between 0.24
mag and 3.46 mag in different sources. Fourteen of 29 FSRQs display
a bluer-when-brighter trend. However, only one source exhibits a
redder-when-brighter trend, which implies that this behavior is rare
in our FSRQ sample. In this source, the thermal emission from the
accretion disk may be responsible for the redder-when-brighter
trend.}
   {}


   \maketitle
%

\section{Introduction}

Blazars, including BL Lac objects and flat-spectrum radio quasars
(FSRQs), are the most extreme class of active galactic nuclei
(AGNs), characterized by strong and rapid variability, high
polarization, and apparent superluminal motion. These extreme
properties are generally interpreted as a consequence of non-thermal
emission from a relativistic jet oriented close to the line of
sight. As such, they represent a fortuitous natural laboratory with
which to study the physical properties of jets, and, ultimately, the
mechanisms of energy extraction from central supermassive black
holes.

While the jet nonthermal emission is generally thought to be the
most dominant type in FSRQs, this is not always true. Chen, Gu \&
Cao (2009a) found that thermal emission can be the most significant
at least in optical bands for some FSRQs selected from the SDSS DR3
quasar catalogue. The emission of FSRQs is indeed a mixture of
thermal emission from accretion disk and jet nonthermal emission. As
thermal emission generally does not vary much as clearly evident in
radio-quiet AGNs (e.g. Ai et al. 2010), any variations in the
emission of FSRQs could be dominated by variations in the jet
nonthermal component. Consequently, the ratio of thermal to
nonthermal emission varies at different source flux states. This may
cause variations in the spectral shape where, for example, the color
becomes redder when the source is brighter as shown in Gu et al.
(2006). In principle, the details of spectral variability may
largely depend on the ratio of thermal to nonthermal emission, and
the characteristics of variations in each component.

The color/spectral behavior can be used to explore the emission
mechanisms in blazars, which remain unclear. The bluer-when-brighter
trend (BWB) is commonly observed in blazars, especially BL Lac
objets (e.g. Ghisellini et al. 1997; Fan et al. 1998; Massaro et al.
1998; Ghosh et al. 2000; Clements \& Carini 2001; Raiteri et al.
2001; Villata et al. 2002; Vagnetti et al. 2003; Wu et al 2005,
2007; Rani et al. 2010), which can usually be explained with
shock-in-jet models. However, the opposite trend, i.e.
redder-when-brighter trend (RWB) has also been found (e.g. Ghosh et
al. 2000; Gu et al. 2006; Rani et al. 2010). The several examples
are 3C 446 (Miller 1981), PKS 0735+178 (Ghosh et al. 2000), PKS
0736+017 (Clements et al. 2003; Ram\'{i}rez et al. 2004), PKS
0420-014 and 3C 454.3 (Gu et al. 2006), and PKS 1622-297 and CTA 102
(Osterman Meyer et al. 2008, 2009). The situation even may be more
complicated because no clear color trend has been reported in some
cases (e.g. Ghosh et al. 2000; Stalin et al. 2006; B\"{o}ttcher et
al. 2007; Poon et al. 2009). Although there have been many studies
of the color behavior of FSRQs, not many FSRQs have been found to
display a redder-when-brighter trend. Moreover, while BL Lac objects
display in general a bluer-when-brighter trend, it is unclear
whether a redder-when-brighter trend is generally present in FSRQs
(Gu et al. 2006; Hu et al. 2006; Rani et al. 2010). In this work, we
investigate the spectral variability of FSRQs, to help us establish
how common the redder-when-brighter trend is in FSRQs.

The layout of this paper is as follows: in Section 2, we describe
the source sample; the variability results are outlined in Section
3; Section 4 includes the discussion; and in the last section, we
draw our conclusions. The cosmological parameters $H_{\rm 0}=70\rm~
km~ s^{-1}~ Mpc^{-1}$, $\Omega_{\rm m}=0.3$, and
$\Omega_{\Lambda}=0.7$ are used throughout the paper, and the
spectral index ¦Á is defined as $f_{\nu}\propto\nu^{-\alpha}$, where
$f_{\nu}$ is the flux density at frequency $\nu$.


\section{Sample selection}

\subsection{Quasars in Stripe 82 region}

Our initial quasar sample was selected as those quasars both in the
SDSS DR7 quasar catalogue (Schneider et al. 2010) and Stripe 82
region. The SDSS DR7 quasar catalogue consists of 105,783
spectroscopically confirmed quasars with luminosities brighter than
$M_{i}=-22.0$, with at least one emission line having a full width
at half-maximum (FWHM) larger than 1000 $\rm km~ s^{-1}$ and highly
reliable redshifts. The sky coverage of the sample is about 9380
$\rm deg^2$ and the redshifts range from 0.065 to 5.46. The
five-band $(u,~ g,~ r,~ i,~ z)$ magnitudes have typical errors of
about 0.03 mag. The spectra cover the wavelength range from 3800 to
9200 $\rm \AA$ with a resolution of $\simeq2000$ (see Schneider et
al. 2010 for details). The Stripe 82 region, i.e. right ascension
$\alpha = 20^{\rm h} - 4^{\rm h}$ and declination
$\delta=-1^{\circ}.25 - +1^{\circ}.25$, was repeatedly scanned
during the SDSS-I phase (2000 - 2005) under generally photometric
conditions, and the data are well calibrated (Lupton et al. 2002).
This region was also scanned repeatedly over the course of three
3-month campaigns in three successive years in 2005 - 2007 known as
the SDSS Supernova Survey (SN survey). The multi-epoch photometric
observations therefore enable us to investigate the optical
variability of the selected quasars.

\subsection{Cross-correlation with radio catalogues}

In this paper, we define a quasar to be a FSRQ according to its
radio spectral index. Therefore, we cross-correlate the initial
quasar sample with the Faint Images of the Radio Sky at Twenty
centimeters (FIRST) 1.4-GHz radio catalogue (Becker, White \&
Helfand 1995), the Green Bank 6-cm (GB6) survey at 4.85 GHz radio
catalogue (Gregory et al. 1996), and the Parkes-MIT-NRAO (PMN) radio
continuum survey at 4.85 GHz (Griffith \& Wright, 1993), as well as
for sources with $\delta<0^{\circ}$. The FIRST survey used the Very
Large Array (VLA) to observe the sky at 20 cm (1.4 GHz) with a beam
size of 5.4 arcsec. FIRST was designed to cover the same region of
the sky as the SDSS, and observed 9000 $\rm deg^2$ at the north
Galactic cap and a smaller $2^{\circ}.5$ wide strip along the
celestial equator. It is 95 per cent complete to 2 mJy and 80 per
cent complete to the survey limit of 1mJy. The survey contains over
800,000 unique sources, with an astrometric uncertainty of $\lesssim
1$ arcsec.

The GB6 survey at 4.85 GHz was executed with the 91-m Green Bank
telescope in 1986 November and 1987 October. Data from both epochs
were assembled into a survey covering the
$0^{\circ}<\delta<75^{\circ}$ sky down to a limiting flux of 18 mJy,
with 3.5 arcmin resolution. GB6 contains over 75,000 sources, and
has a positional uncertainty of about 10 arcsec at the bright end
and about 50 arcsec for faint sources (Kimball \& Ivezi\'{c} 2008).
The PMN surveys were made using the Parkes 64-m radio telescope at a
frequency of 4850 MHz with the NRAO multibeam receiver mounted at
the prime focus (Griffith \& Wright 1993). The surveys had a spatial
resolution of approximately $4'.2$ FWHM and were made for the
southern sky between declinations of $-87^{\circ}$ and
$+10^{\circ}$, and all right ascensions during June and November in
1990. The positional accuracy is close to 10 arcsec in each
coordinate. The survey was divided into four declination bands. One
of these four is the equatorial survey
($-9^{\circ}.5<\delta<+10^{\circ}.0$) covering 1.90 sr, which
contains 11,774 sources to a flux limit of 40 mJy and largely
overlaps the GB6 survey in the declination range from $0^{\circ}$ to
$+10^{\circ}$ (Griffith et al. 1995).

The initial quasar sample was first cross-correlated between the
SDSS quasar positions and the FIRST catalogue to within 2 arcsec
(see e.g. Ivezi\'{c} et al. 2002; Lu et al. 2007). The resulting
sample of SDSS quasar positions was then cross-correlated with both
the GB6 and PMN equatorial catalogues to within 1 arcmin (e.g.
Kimball \& Ivezi\'{c} 2008). Owing to the different spatial
resolutions of FIRST, GB6, and PMN, multiple FIRST counterparts were
found to within 1 arcmin for some quasars, although there is only
single GB6 and/or PMN counterparts existed. The flux density of the
closest FIRST counterparts may be dominant among multiple
counterparts in some sources, although not in others. In this paper,
we conservatively selected the sources with single FIRST
counterparts to within 1 arcmin of the SDSS positions. This
selection criterion, for example, helps avoid the contamination of
genuine steep-spectrum radio quasars (SSRQs). These SSRQs may appear
as multi-FIRST components, but be identified as FSRQs if using the
flux density of the closest FIRST component, which is actually much
smaller than that of radio lobes. The optical variability of quasars
with multiple FIRST counterparts will be presented in a forthcoming
paper.

A sample of 37 quasars was constructed from the above
cross-correlations. The radio spectral index $\alpha$ was then
calculated between the FIRST 1.4 and either or both of the GB6 and
PMN 4.85 GHz. Finally, 32 quasars were conventionally defined as
FSRQs, with $\alpha<0.5$, and the remaining five quasars are SSRQs
with $\alpha>0.5$. When counterparts were found in both the GB6 and
PMN surveys, the spectral indices were consistent with each other.

\subsection{Sample}

\begin{figure}
   \centering
   \includegraphics[width=0.45\textwidth]{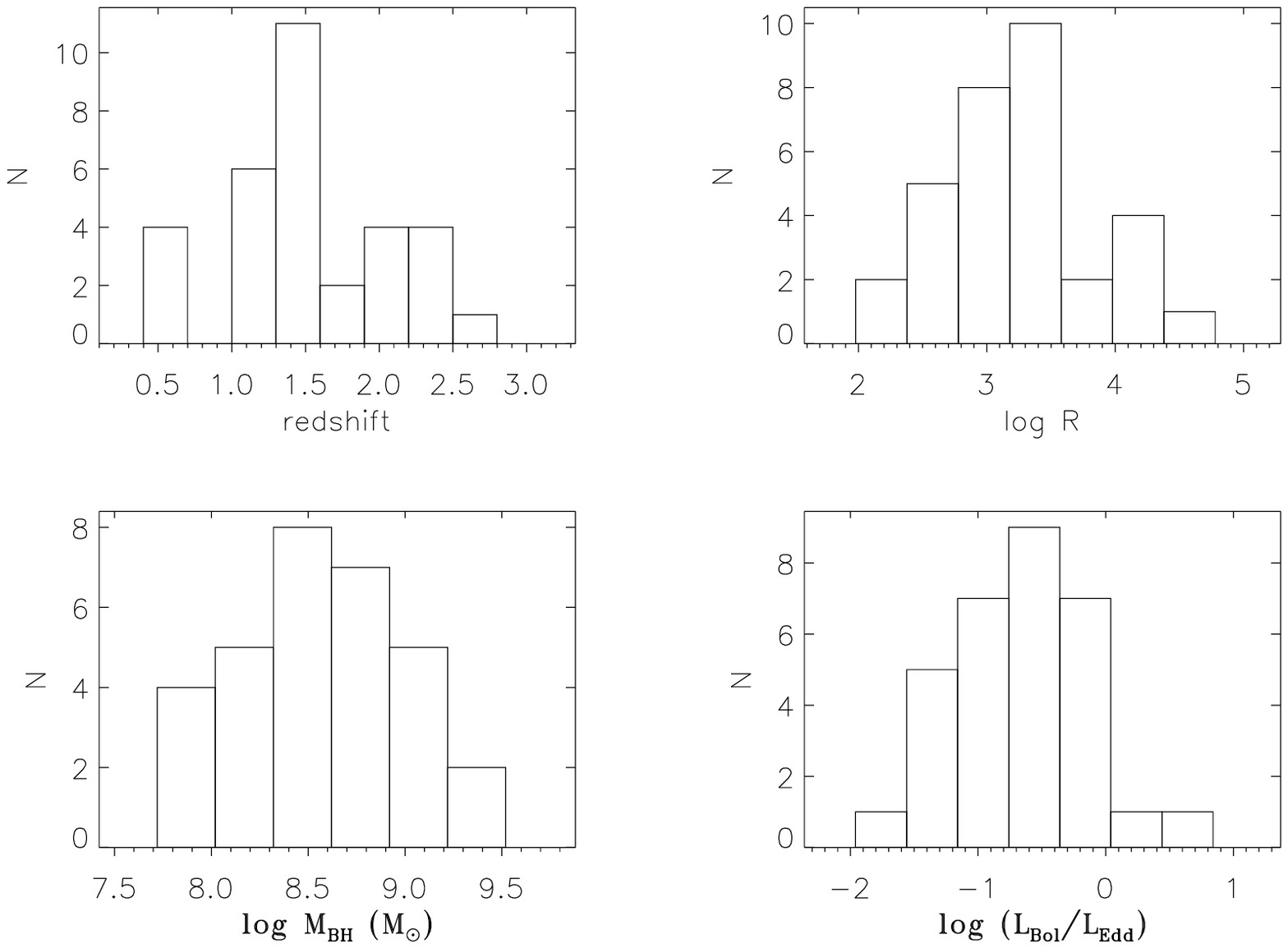}
   \caption{The histogram of sample parameters: redshift (upper left), radio loudness (upper right),
   black hole mass (lower left), and the Eddington ratio (lower right). SDSS J222646.53+005211.3 is not
   included in the histogram of black hole mass and Eddington ratio because of the lack of broad emission-line measurements.}
              \label{hist}%
    \end{figure}

Our final sample of 32 FSRQs is listed in Table \ref{table_source},
which provides the source redshift, radio loudness, radio spectral
index, black hole mass, disc bolometric luminosity, and the
Eddington ratio, and the similar data of five SSRQs are listed for
comparison. The distribution of redshift, radio loudness, black hole
mass, and the Eddington ratio are shown in Fig. \ref{hist}. The
source redshift is taken directly from the SDSS DR7 quasar
catalogue, which covers $0.4<z<2.7$. The radio loudness is from Shen
et al. (2010), which ranges from $\rm log~ \it R\rm =1.97$ (SDSS
J005205.56+003538.1) to $\rm log ~\it R\rm =4.48$ (SDSS
J222646.53+005211.3). However, the radio loudness was calculated as
$R=f_{\rm 6cm}/f_{2500}$, where $f_{\rm 6cm}$ and $f_{2500}$ are the
flux density at rest-frame 6 cm and $2500\AA$, respectively (see
Shen et al. 2010 for more details). Among 32 FSRQs, 14 sources have
an inverted spectral index between 1.4 and 4.85 GHz with $\alpha<0$.

Black hole masses are estimated from the various empirical relations
in the literature by using the luminosity and FWHM of broad $\rm
H\beta$, Mg II, and C IV lines, i.e., Vestergaard \& Peterson (2006)
for $\rm H\beta$, and Kong et al. (2006) for Mg II and C IV. The
luminosity and FWHM of broad $\rm H\beta$, Mg II, and C IV lines are
adopted from the measurements in Shen et al. (2010). The BLR
luminosity $L_{\rm BLR}$ is derived following Celotti, Padovani \&
Ghisellini (1997) by scaling the strong broad emission lines $\rm
H\beta$, Mg II, and C IV to the quasar template spectrum of Francis
et al. (1991), in which $\rm Ly \alpha$ is used as a flux reference
of 100. By adding the contribution of $\rm H\alpha$ with a value of
77, the total relative BLR flux is 555.77, which consists of $\rm
H\beta$ at 22, Mg II at 34 and C IV at 63 (Francis et al. 1991;
Celotti et al. 1997). From the BLR luminosity, we estimate the disc
bolometric luminosity as $L_{\rm Bol} = 10 L_{\rm BLR}$ (Netzer
1990). The black hole mass ranges from $10^{7.7} M_{\odot}$ to
$10^{9.4} M_{\odot}$, most of sources being in $10^{8.5} - 10^{9.0}
M_{\odot}$ range, which is typical of blazars (Ghisellini et al.
2010). The Eddington ratio $\rm log~ \it L_{\rm Bol}/L_{\rm Edd}$
ranges from -1.96 to 0.57, with most sources being in the -1.0 - 0.0
range, which is indicative of a standard thin disk (Shakura \&
Sunyaev 1973) being present in all FSRQs.

\section{Results}

The SDSS DR7 CAS contains the Stripe82 database, containing all
imaging from SDSS stripe 82 along the celestial equator at the
southern Galactic cap. It includes a total of 303 runs, covering any
given piece of the close to 270 $\rm deg^2$ area approximately 80
times. Only about one-quarter of the stripe 82 scans were obtained
in photometric conditions, the remainder being taken under variable
clouds and often poorer than normal seeing. For the runs that were
non-photometric, an approximate calibration, using the photometric
frames as reference, was derived and made available in the CAS
Stripe82 database. In this work, we directly use the
point-spread-function magnitudes in the CAS Stripe82 database from
the photometric data obtained during the SDSS-I phase from data
release 7 (DR7; Abazajian et al. 2009) and the SN survey during 2005
- 2007. The typical measurement error in magnitude is about 0.03
mag.

Among 32 FSRQs, we select the sources classified as point sources in
all observational runs. Only data with good measurements
(high-quality photometry) are selected following the recommendations
in the SDSS
instructions\footnote{http://www.sdss.org/dr7/products/catalogs/flags.html}.
Moreover, we insist on the $ugriz$ magnitude being brighter than the
magnitude limit in each band, i.e. 22.0, 22.2, 22.2, 21.3, 20.5 at
$u$, $g$, $r$, $i$, $z$, respectively. The data taken at cloudy
conditions are also excluded. We calculate the spectral index from
the linear fit to the $\rm log~\it f_{\nu} - \rm log ~\nu$ relation
after applying an extinction correction to the $ugriz$ flux density
and taking the flux error into account. In most cases, the linear
fit gave good fits. Each cycle of $ugriz$ photometry was usually
completed within five minutes, i.e. quasi-simultaneously, therefore,
the spectral index calculation will not be seriously influenced by
any source variations.

\subsection{Variability}

Among 32 FSRQs, three sources were excluded from our analysis for
various reasons. SDSS J015243.14$+$002039.6 is classified as a
galaxy in the SDSS pipeline, although it is classified as a quasar
in the SDSS DR7 quasar catalogue. The images of SDSS
J215349.75$+$003119.5 are usually saturated because of a nearby
bright object. Owing to the low quality of the $u$ band images, SDSS
J222646.53$+$005211.3 was also excluded because we aim to derive the
spectral index using only high quality $ugriz$ photometric data.

\begin{figure}
   \centering
   \includegraphics[width=0.45\textwidth]{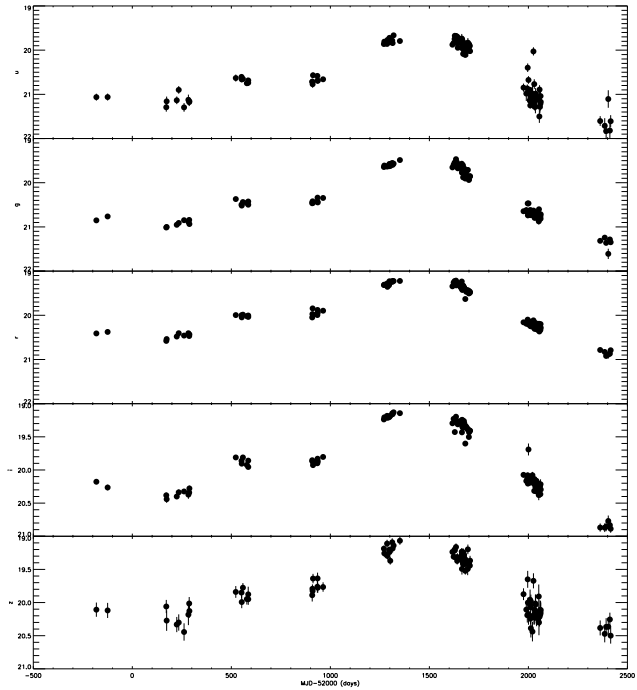}
   \caption{The $ugriz$ band light curve of SDSS J025515.09+003740.5 (from top to bottom).}
              \label{J02lc}%
    \end{figure}

All remaining 29 FSRQs show large amplitude variations with overall
variations in r band $\Delta r=0.24 - 3.46$ mag (see Table
\ref{table_source}), which is much larger than that of radio quiet
AGNs, 0.05 - 0.3 mag (e.g. Ai et al. 2010), but typical of blazars
(e.g. Gu et al. 2006). There are four sources with $\Delta r>1$ mag,
i.e. SDSS J001130.400+005751.80 - $\Delta r=3.46$ mag, SDSS
J023105.597+000843.61 - $\Delta r=1.02$ mag, SDSS
J025515.096+003740.55 - $\Delta r=1.70$ mag, and SDSS
J235936.817-003112.78 - $\Delta r=1.20$ mag. In general, the
variations in different bands show similar trends, as can be seen
from the light curve of SDSS J025515.096+003740.55 for example in
Fig. \ref{J02lc}. This source has been observed every year from 2000
to 2007. From Fig. \ref{J02lc}, the source remained stable in 2000 -
2001, gradually brightening after that. It probably reached its
brightest state between the 2004 and 2005 observational runs, after
which the source gradually decreased in brightness until reaching
its faintest magnitude in 2007. The fluctuations can also be seen in
each observational session, for example, there having been about 0.5
mag of amplitude variations in the observational run of 2005.

\subsection{Spectral index \& brightness relation}

The correlation between the spectral index $\alpha_{\nu}$ and PSF
$r$ magnitude was checked for all sources using the Spearman rank
correlation analysis method. We found that 15 of 29 FSRQs show a
significant correlation at a confidence level of $>99\%$, of which
14 FSRQs show positive correlations, and only one FSRQ (SDSS
J001130.40$+$005751.7, see next section for details) shows a
negative correlation (see Table 1). In contrast, two of the five
SSRQs have significant positive correlations at a confidence level
of $>99\%$ (see Table 1).

\begin{figure}
   \centering
   \includegraphics[width=0.45\textwidth]{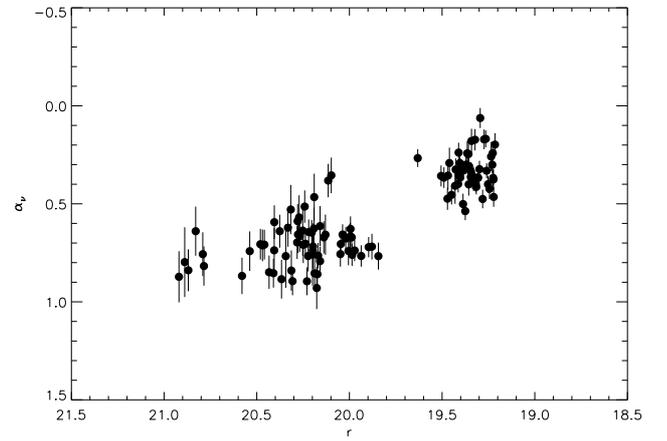}
   \caption{The relationship between the spectral index and the PSF magnitude
   at r band for SDSS J025515.09+003740.5. A significant positive correlation is
   present, which implies a bluer-when-brighter trend.}
              \label{J02ra}%
    \end{figure}

As an example, the correlation between $\alpha_{\nu}$ and $r$
magnitude is shown in Fig. \ref{J02ra} for SDSS
J025515.096+003740.55. A significant positive correlation can be
clearly seen. The Spearman correlation analysis gives a correlation
coefficient of $r_{s}=0.762$ at a confidence level of $\gg99.99\%$.
This positive correlation means that the source spectra becomes
flatter as the source brightens, in other words, the source follows
a bluer-when-brighter trend. We noted that the spectral indices are
all smaller than 1.0, implying a rising SED in the $ugriz$ region,
more strictly, in $1755 - 4415~ \AA$ wavelength region in the source
rest-frame (source redshift $z=1.0228$). Since the overall variation
amplitude $\Delta r=1.70$ mag is much larger than the typical values
of radio-quiet AGNs (e.g. Ai et al. 2010), the variations are most
likely to originate from jet nonthermal emissions. Therefore, the
bluer-when-brighter trend implies that the synchrotron peak
frequency likely increases as the source brightens. Assuming that
the source SED in the $ugriz$ region closely follows the overall SED
of synchrotron emission, $\alpha_{\nu}<1.0$ means that the
synchrotron peak frequency may likely be $\nu_{\rm peak}>10^{15.23}$
Hz (i.e. correspond to wavelengths shorter than $1755 ~\AA$). This
would indicate that this source should be classified as a
high-synchrotron-peaked FSRQ which however is not common in FSRQs
(Abdo et al. 2010, see also Chen et al. 2009a). Interestingly, we
found that 17 of 29 FSRQs have $\alpha_{\nu}<1.0$ all the time
during variability (see Col. 9 of Table \ref{table_source}), thus
likely have a high synchrotron peak frequency if their $ugriz$ SED
closely follows the overall synchrotron SED. Nevertheless, the firm
SED classification should be verified by simultaneous multi-band
data with a much wider SED coverage.

\subsection{SDSS J001130.40$+$005751.7}

As shown in the previous section, SDSS J001130.40$+$005751.7 is the
only FSRQ displaying a negative correlation between the spectral
index and $r$ magnitude. The negative correlation shows that the
source spectrum becomes steeper when the source is brighter, i.e.
that there is a redder-when-brighter trend. This source is a known
FSRQ, which is included in the Radio Optical X-ray ASDC (ROXA)
blazar sample (Turriziani, Cavazzuti \& Giommi 2007). It appears as
a compact radio source in the NVSS and FIRST maps, with a flux
density of $f_{\rm FIRST}=156.07$ mJy, and $f_{\rm NVSS}=167.2$ mJy
in FIRST and NVSS, respectively. The spectral index calculated for
FIRST 1.4 GHz and PMN 4.85 GHz flux density of $f_{\rm PMN}=170$ mJy
is $\alpha=-0.07$, while it is $\alpha=0.13$ when the GB6 4.85 GHz
flux density of $f_{\rm GB6}=132.76$ mJy is used. It is in the fifth
VLBA calibrator survey (VCS5: Kovalev et al. 2007), and is compact
in the 8.6 GHz VLBA map, with an integrated flux density of $f_{\rm
int} = 90$ mJy, and a flux density for the unresolved component of
$f_{\rm unre} = 70$ mJy. It is also included in the Combined Radio
All-sky Targeted Eight GHz Survey (CRATES: Healey et al. 2007) with
$f_{\rm 8.4 GHz} = 278.7$ mJy from VLA maps.

\begin{figure}
   \centering
   \includegraphics[width=0.45\textwidth]{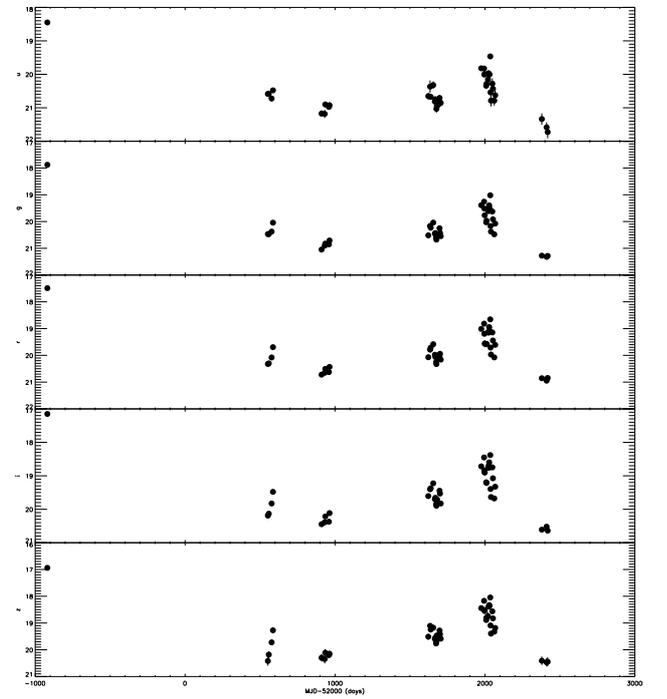}
   \caption{The $ugriz$ band light curve of SDSS J001130.40$+$005751.7 (from top to bottom). }
              \label{J00lc}%
    \end{figure}

\begin{figure}
   \centering
   \includegraphics[width=0.45\textwidth]{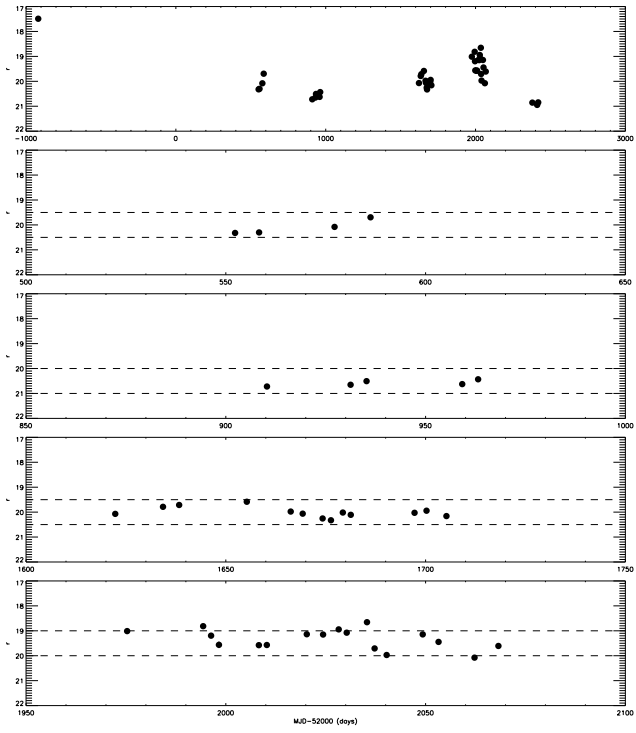}
   \caption{The r band light curve of SDSS J001130.40$+$005751.7. The light curves of the second
   to fifth observational sessions are zoomed in the second to fifth panels from
   up to bottom. The dashed lines represent 1 mag variations.}
              \label{J00lcr}%
    \end{figure}

The light curves of SDSS J001130.40$+$005751.7 at u, g, r, i, and z
bands are shown in Fig. \ref{J00lc}. The variations exhibits similar
trends in all bands, over nine years from 1998 to 2007. The overall
variation in r band is 3.46 mag, which is shown in Fig.
\ref{J00lcr}. In the observational sessions, the source was
brightest at MJD=51081 with $r=17.49$, while faintest at MJD=54411
with $r=20.95$. This source seems to vary all the time, with about 1
mag fluctuations in most observational sessions (see Fig.
\ref{J00lcr}).

\begin{figure}
   \centering
   \includegraphics[width=0.45\textwidth]{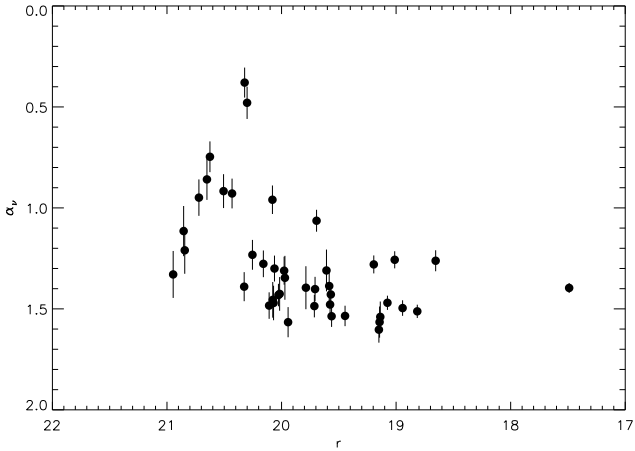}
   \caption{The relationship between the spectral index and the PSF magnitude
   at r band for SDSS J001130.40$+$005751.7. A significant anti-correlation is
   present, which implies a redder-when-brighter trend.}
              \label{J00ra}%
    \end{figure}

The negative correlation between the spectral index $\alpha_{\nu}$
and $r$ magnitude is shown in Fig. \ref{J00ra}. The Spearman
correlation analysis shows a significant negative correlation with a
correlation coefficient of $r_{\rm s}=-0.606$ at confidence level of
$>99.99\%$. This source is included in the first Fermi Large Area
Telescope AGN catalogue with a photon index of $2.51\pm0.15$ and
classified as a low-synchrotron-peaked FSRQs with $\nu_{\rm
peak}<10^{14}$ Hz (Abdo et al. 2010). With the source redshift
$z=1.4934$, SDSS $ugriz$ wavebands correspond to the wavelength
range of $1424 - 3582 \AA$ in the source rest-frame, which is
therefore likely at the falling part of synchrotron SED. For a
sample of 17 radio-quiet AGNs, Shang et al. (2005) show that the
spectral break occurs at around $1100~ \AA$ for most objects, which
is thought to be closely related to the big blue bump. If this
spectral break also exists in SDSS J001130.40$+$005751.7, we would
expect to observe the rising part of accretion disk thermal emission
when it dominates over the nonthermal emission. While the spectral
index and brightness generally follows an anti-correlation, we note
that it is more complex than a simple anti-correlation, as shown in
Fig. \ref{J00ra}. While the rising optical spectral
($\alpha_{\nu}<1.0$) is indeed present at low flux state, the
declining spectra ($\alpha_{\nu}>1.0$) are also clearly visible at
low fluxes. In contrast, the optical spectra decline
($\alpha_{\nu}>1.0$) at high flux states, which implies that
nonthermal low-peak-frequency synchrotron emission then provides the
greatest contribution.


\section{Discussions}

The bluer-when-brighter trend is commonly observed in blazars, as it
has been found by different investigators that the amplitude of the
variations is systematically larger at higher frequencies,
suggesting that the spectrum becomes steeper when the source
brightness decreases, and flatter when it increases (Racine 1970;
Gear, Robson \& Brown 1986; Ghisellini et al. 1997; Massaro et al.
1998). Investigations of spectral variability have also detected
this general trend (D'Amicis et al. 2002; Vagnetti et al. 2003;
Fiorucci, Ciprini \& Tosti 2004). From our investigations, the
variations in the spectral index of 14 of 29 FSRQs indeed follow
this trend. This common phenomenon has different possible
explanations (Fiorucci et al. 2004). It may indicate the presence of
two components that contribute to the overall emission in the
optical region, one variable (with a flatter slope $\rm
\alpha_{var}$, where $f_{\nu}\propto\nu^{-\alpha}$), and the other
stable (with $\rm \alpha_{const}>\alpha_{var}$). Alternatively, it
may be interpreted using a one-component synchrotron model: the more
intense the energy release, the higher the particle's energy
(Fiorucci et al. 2004). Another possible explanation is that the
luminosity increase was caused by the injection of fresh electrons,
with an energy distribution harder than that of the previous,
partially cooled ones (e.g. Kirk, Rieger \& Mastichiadis 1998;
Mastichiadis \& Kirk 2002). Finally, it could be due to Doppler
factor variations in a spectrum slightly deviating from a power law,
e.g. Doppler factor variations in a `convex' spectrum (Villata et
al. 2004).

However, some evidence that the amplitudes of variations are not
systematically larger at high frequencies has been found on several
occasions (see, for example: Malkan \& Moore 1986; Brown et al.
1989; Massaro et al. 1998; Ghosh et al. 2000; Clements et al. 2003;
Ram\'{i}rez et al. 2004). Based on their results, Ghosh et al.
(2000) suggested that it may be incorrect to generalize by saying
that the amplitude of the variation in blazars is systematically
larger at higher frequency. They found that the spectral slope of AO
0235+164 remained almost constant when its brightness increased. In
S5 0716+714, while the long-term spectral change is evidently BWB,
both BWB chromatism and achromatism were found in both bright and
dim source flux states among the microvariability nights (Poon et
al. 2009). The authors claimed that the BWB behavior can be
explained by the shock-in-jet model, while the achromatic trend may
be due to geometric effects. Interestingly, we did not find any
significant correlations between spectral index and brightness for
14 FSRQs in our sample.

The redder-when-brighter trend has also been found in several cases
(e.g. Gu et al. 2006, Ram\'{i}rez et al. 2004). The competition
between the thermal accretion disk and synchrotron emission was used
to explain the RWB trend in FSRQs (Gu et al. 2006). The optical
emission of FSRQs is contaminated by thermal emission from the
accretion disk. Since FSRQs are usually low-synchrotron-peaked
sources with $\nu_{\rm peak}<10^{14}$ Hz (Abdo et al. 2010), the big
blue bump of accretion disk emission will flatten the spectral slope
in the optical region, resulting in a flatter composite spectrum
than the non-thermal component. When the object brightens, the
non-thermal component has a more dominant contribution to the total
flux, and the composite spectrum steepens. The existence of thermal
accretion disk emission is supported by the UV excess in 3C 454.3
during lower flux states (Raiteri et al. 2007). The thermal blue
bump has also been observed in 3C 345 (Webb et al. 1994).
Interestingly, the emergence of accretion disk emission was detected
in NLS1s PMN J0948+0022 as its synchrotron emission dropped (Abdo et
al. 2009). While the spectral behaviors are not homogenous across
our FSRQs sample, it may also be inhomogenous even for a single
source. From the four-year five-waveband monitoring, Wu et al.
(2010) found both a bluer-when-brighter and redder-when-brighter
trend in FSRQ 3C 345 when using a different color index. The RWB
trend can be explained by the less variable thermal components.
Moreover, the authors argued that the color behaviors of FSRQs are
linked not only to their emission process, but also to the redshift
and strengths of the less variable thermal components (Wu et al.
2010). Hence, it is most likely that the spectral behaviors of FSRQs
depend on the position of the synchrotron peak frequency, the
sampled optical wavelength range in the source rest frame (which
depends on the redshift), and the positions of thermal blue bump and
its strength compared to the jet emission. We note that FSRQs are
supposed to display a redder-when-brighter trend, for example, in at
least two of three sources, i.e., 3C 454.3 and PKS 0420-014 (Gu et
al. 2006), and four of the six sources, PKS 0420-014, 4C 29.45, PKS
1510-089, and 3C 454.3 (Rani et al. 2010), all of which are known
low-synchrotron-peaked FSRQs. In contrast, our results imply that
the redder-when-brighter trend may be rarer for FSRQs, at least for
our present sample. However, the bluer-when-brighter trend is more
common.

\begin{figure}
   \centering
   \includegraphics[width=0.45\textwidth]{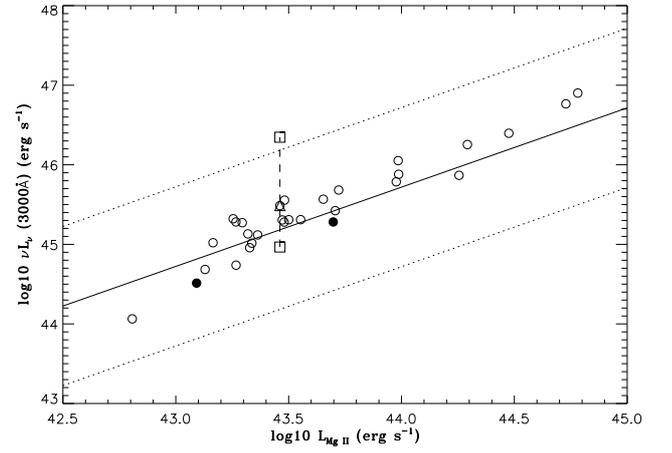}
   \caption{The plot of broad Mg II line luminosity and continuum luminosity
   at 3000 $\AA$. The solid line is the OLS bisector linear fit to radio-quiet
   AGNs in Kong et al. (2006), $\lambda L_{\rm \lambda~ 3000 \AA}=78.5~ L^{0.996}_{\rm Mg~ II}$. The open
circles are FSRQs, solid circles are SSRQs, and open
   triangle is SDSS J001130.40$+$005751.7. The vertical dashed line represents the luminosity
   range at 3000 $\AA$ assuming its variation is identical to that in the $r$ band, which is connected by
   two rectangles. The dotted lines
   show the one order of magnitude deviation of continuum luminosity
   at 3000 $\AA$ from the solid line.}
              \label{mgcon}%
    \end{figure}

Radio-loud quasars, particularly FSRQs, differ from radio-quiet AGNs
in terms of their contribution of the nonthermal jet emission in
optical continuum emission, in addition to the thermal emission from
the accretion disk. In particular, jet emission can be dominant
because of the beaming effect when the jet is moving towards us with
a small viewing angle. To inspect the continuum emission in our
FSRQs, we plot in Fig. \ref{mgcon} the relationship between the
continuum luminosity at 3000 $\rm \AA$ and broad Mg II line
luminosity for 28 FSRQs, both of which are adopted from Shen et al.
(2010). For comparison, the ordinary least-square bisector linear
fit $\lambda L_{\rm \lambda~ 3000 \AA}=78.5~ L^{0.996}_{\rm Mg~ II}$
for a sample of radio-quiet AGNs in Kong et al. (2006) is also
plotted with the solid line (see also Chen et al. 2009b). It can be
seen that the $3000~\rm \AA$ luminosity of most sources lies above
the $\lambda L_{\lambda 3000~\AA} - L_{\rm Mg II}$ relation of
radio-quiet AGNs. However, the deviation of $\lambda L_{\lambda
3000~\AA}$ is within 0.5 dex. This excess is most likely to be
caused by the jet nonthermal emission. The spectra of SDSS
J001130.40$+$005751.7 at MJD=51793 was used to measure the broad
emission lines and continuum (Shen et al. 2010), from which the
corresponding r magnitude is about 19.6. If the luminosity at 3000
$\rm \AA$ of SDSS J001130.40$+$005751.7 follows the variation
amplitude of the r magnitude, we can estimate its highest and lowest
values, which are indicated by two rectangles assuming a same Mg II
luminosity. Therefore, in the highest flux state, the continuum
emission could be dominated completely by the nonthermal emission by
more than one order of magnitude (see Fig. \ref{mgcon}). However, in
the lowest flux state, the thermal emission is likely to be
responsible
for the continuum emission. 

While FSRQs are usually associated with core-dominated radio
quasars, SSRQs are generally related to lobe-dominated ones, usually
with two large-scale optically thin radio lobes. Usually, the
beaming effect is not severe in SSRQs because of the relatively
large viewing angle. Therefore, the optical continuum of SSRQs could
be dominated by the thermal emission. In Fig. \ref{mgcon}, two SSRQs
with available Mg II measurements all lie close to but below the
solid line for radio quiet AGNs. This likely implies that the
thermal emission from accretion disk is indeed dominated in the
optical continuum at the epoch of SDSS spectra. Interestingly, two
of five SSRQs display a BWB trend (see Table \ref{table_source}).
The BWB trend has also been found in radio quiet AGNs, and the
variability was found to be anti-correlated with the Eddington ratio
(Ai et al. 2010). In radio quiet AGNs, the optical variability could
be due to the variation in the accretion process, for example, the
variation in the accretion rate (e.g. Li \& Cao 2008), since the jet
is either weak or absent. These explanations could also be used to
explain the variability of SSRQs if the optical emission is indeed
mainly from the accretion disk. However, that the variation is
caused by jet nonthermal emission cannot be completely excluded in
some cases because the variability amplitudes are generally larger
that the typical values for radio quiet AGNs (see Table
\ref{table_source}). In extreme cases, the overall amplitude is
$\Delta r=0.84$ mag in SDSS J012401.76+003500.9. It may be more
likely that both thermal and nonthermal emission contribute to the
variability. Further multi-waveband monitoring is needed to help
resolve these uncertainties, especially the spectroscopic
monitoring.

\section{Summary}

We have constructed a sample of 32 FSRQs in the SDSS Stripe 82
region. The variability and the relationship between the spectral
index and brightness were investigated for 29 FSRQs. We found that
all FSRQs show large-amplitude overall variations, e.g. from 0.24 to
3.46 mag at r band. We only found a significant negative correlation
between the spectral index and r magnitude (i.e.
redder-when-brighter) in one FSRQ, SDSS J001130.40$+$005751.7. This
implies that the RWB trend is rare in our FSRQ sample. This behavior
could be explained by the contribution of thermal accretion disk
emission in the optical wavebands. In contrast, a
bluer-when-brighter trend is more common in FSRQs, and found in 14
of the 29 FSRQs studied here.

%

\begin{acknowledgements}
We thank the anonymous referee for helpful comments and suggestions.
MFG thanks R. Antonucci, S. H\"{o}nig, J. Wu and Z. Chen for useful
discussions. This work is supported by the National Science
Foundation of China (grants 10703009, 10821302, 10833002, 10978009,
11033007 and 11073039), and by the 973 Program (No. 2009CB824800).
Funding for the SDSS and SDSS-II was provided by the Alfred P. Sloan
Foundation, the Participating Institutions, the National Science
Foundation, the U.S. Department of Energy, the National Aeronautics
and Space Administration, the Japanese Monbukagakusho, the Max
Planck Society, and the Higher Education Funding Council for
England. The SDSS Web site is http://www.sdss.org/.
\end{acknowledgements}

\clearpage

\newpage

\begin{table}
\caption{\label{table_source}Source list: Col. 1 - SDSS source name;
Col. 2 - redshift; Col. 3 - radio loudness; Col. 4 - the spectral
index between 1.4 and 4.85 GHz, of which PMN flux density is used
for those labeled with $^{a}$, otherwise GB6 one is used; Col. 5 -
black hole mass in unit of solar mass; Col. 6 - disc bolometric
luminosity in unit of $\rm erg~s^{-1}$; Col. 7 - the Eddington ratio
$l=L_{\rm BOL}/L_{\rm EDD}$; Col. 8 - overall variation in r band;
Col. 9 - the range of variation in the spectral index
$\alpha_{\nu}$; Cols. (10-11) - the Spearman correlation coefficient
and probability level,
respectively.
} \centering
\begin{tabular}{lcccccccccc}
\hline\hline
SDSS source   &    $z$   & log 10 (R) & $\alpha_{\rm r}$ & log 10 ($M_{\rm BH}$) & log 10 ($L_{\rm BOL}$) & log 10 ($l$) & $\Delta~ r$ & $\Delta~ \alpha_{\nu}$ & $r_{\rm s}$ & prob. \\
 & & & & ($\rm M_{\odot}$) & ($\rm erg~s^{-1}$) &  & (mag) & & \\ 
(1)&(2)&(3)&(4)&(5)&(6)&(7)&(8)&(9)&(10)&(11)\\
\hline
J001130.40$+$005751.7   &  1.4934    &  3.42  &    0.13        &   8.59   &    45.67   &   -1.02  &   3.45   &  0.38,1.60    &   -0.606   &  0.16e-04  \\
J001611.08$-$001512.4   &  1.5759    &  4.26  &    0.32$^{a}$  &   8.40   &    45.91   &   -0.60  &   0.55   &  0.86,1.74    &    0.250   &  0.11      \\
J002509.66$-$004031.0   &  2.2086    &  3.21  &    0.25$^{a}$  &   8.54   &    46.41   &   -0.24  &   0.37   &  0.28,0.71    &    0.310   &  0.16e-01  \\
J003443.92$-$005413.0   &  0.6561    &  3.03  &   -0.63$^{a}$  &   8.84   &    45.41   &   -1.54  &   0.50   &  -0.19,0.52   &   -0.167   &  0.15      \\
J003512.90$+$010430.6   &  2.2603    &  3.45  &    0.49        &   8.77   &    45.85   &   -1.04  &   0.34   &  0.71,1.35    &    0.209   &  0.29      \\
J004332.71$+$002459.8   &  1.1277    &  3.21  &    0.09        &   8.02   &    45.47   &   -0.67  &   0.47   &  -0.28,1.24   &    0.452   &  0.20e-07  \\
J004819.12$+$001457.1   &  1.5450    &  3.45  &   -0.06$^{a}$  &   8.36   &    45.93   &   -0.55  &   0.52   &  0.02,0.90    &    0.769   &  0.29e-13  \\
J005205.56$+$003538.1   &  0.3993    &  1.97  &    0.01        &   8.51   &    45.97   &   -0.65  &   0.44   &  0.83,1.49    &   -0.100   &  0.47      \\
J005716.99$-$002433.2   &  2.7316    &  2.93  &   -0.42$^{a}$  &   9.11   &    46.68   &   -0.55  &   0.49   &  0.17,0.77    &   -0.107   &  0.46      \\
J011129.94$+$003431.3   &  1.3203    &  2.47  &   -0.59        &   8.00   &    45.94   &   -0.18  &   0.54   &  0.11,0.97    &   -0.007   &  0.95      \\
J012401.76$+$003500.9   &  1.8516    &  3.85  &    0.86        &   9.36   &    46.11   &   -1.36  &   0.84   &  -0.52,0.61   &    0.687   &  0.68e-06  \\
J012517.14$-$001828.9   &  2.2780    &  3.44  &    0.59$^{a}$  &   8.50   &    47.13   &    0.52  &   0.29   &  0.22,0.87    &   -0.202   &  0.11      \\
J012528.84$-$000555.9   &  1.0759    &  3.16  &    0.12$^{a}$  &   9.13   &    46.69   &   -0.55  &   0.50   &  -0.59,0.25   &    0.012   &  0.91      \\
J012753.70$+$002516.4   &  2.4566    &  3.46  &    0.42$^{a}$  &   7.82   &    46.51   &    0.57  &   0.62   &  -0.34,0.49   &    0.216   &  0.42e-01  \\
J015243.14$+$002039.6   &  0.5776    &  3.95  &    0.15        &   8.16   &    45.02   &   -1.26  &   ...    &  ...          &    ...     &  ...       \\
J015454.36$-$000723.1   &  1.8287    &  3.52  &    0.32$^{a}$  &   8.46   &    46.47   &   -0.10  &   0.73   &  0.16,0.91    &    0.382   &  0.90e-03  \\
J015832.51$-$004238.2   &  2.6071    &  3.58  &    0.87$^{a}$  &   8.29   &    46.90   &    0.50  &   0.47   &  0.44,1.19    &   -0.207   &  0.28      \\
J021728.62$-$005227.2   &  2.4621    &  3.29  &    0.75$^{a}$  &   8.84   &    46.40   &   -0.55  &   0.35   &  0.45,1.28    &    0.488   &  0.77e-05  \\
J022508.07$+$001707.2   &  0.5270    &  3.81  &    0.86        &   8.88   &    45.19   &   -1.80  &   0.53   &  0.25,0.68    &    0.470   &  0.27e-01  \\
J023105.59$+$000843.5   &  1.3354    &  3.03  &    0.10$^{a}$  &   8.63   &    45.77   &   -0.98  &   1.02   &  0.09,0.76    &    0.663   &  0.10e-07  \\
J025515.09$+$003740.5   &  1.0228    &  2.86  &   -0.78        &   8.73   &    45.55   &   -1.29  &   1.70   &  0.06,0.93    &    0.761   &  0.46e-22  \\
J213638.58$+$004154.2   &  1.9397    &  3.69  &   -0.87        &   9.11   &    47.29   &    0.07  &   0.28   &  0.12,0.52    &    0.581   &  0.10e-06  \\
J215349.75$+$003119.5   &  1.8144    &  3.56  &    0.37        &   8.32   &    46.07   &   -0.36  &   ...    &  ...          &    ...     &  ...       \\
J221001.81$-$001309.7   &  1.0941    &  3.54  &    0.31$^{a}$  &   7.84   &    45.54   &   -0.41  &   0.47   &  0.19,1.07    &    0.476   &  0.33e-02  \\
J221806.67$+$005223.5   &  1.2734    &  2.04  &   -0.23        &   8.69   &    46.51   &   -0.30  &   0.26   &  -0.07,0.33   &    0.656   &  0.87e-09  \\
J222235.84$+$001536.3   &  1.3617    &  3.28  &    0.24$^{a}$  &   8.83   &    45.53   &   -1.41  &   0.65   &  0.20,1.21    &    0.778   &  0.33e-08  \\
J222646.53$+$005211.3   &    2.26    &  4.48  &    0.15        &   ...    &    ...     &    ...   &   ...    &  ...          &    ...     &  ...       \\
J222713.23$-$005041.4   &  1.5063    &  2.62  &   -1.12$^{a}$  &   8.75   &    45.78   &   -1.08  &   0.87   &  0.60,1.89    &    0.766   &  0.15e-13  \\
J222744.58$+$003450.8   &  1.5488    &  2.67  &   -0.57        &   8.11   &    45.64   &   -0.59  &   0.65   &  -0.01,0.91   &    0.581   &  0.46e-06  \\
J224146.37$+$001608.0   &  1.4985    &  3.08  &   -0.06$^{a}$  &   7.72   &    45.51   &   -0.33  &   0.40   &  -0.07,0.78   &    0.761   &  0.77e-08  \\
J231222.36$-$010924.8   &  1.4309    &  4.24  &    0.37$^{a}$  &   8.44   &    45.58   &   -0.98  &   0.28   &  1.44,2.47    &   -0.068   &  0.73      \\
J232037.99$+$003139.7   &  1.8993    &  2.99  &   -0.18        &   9.07   &    46.55   &   -0.64  &   0.47   &  0.23,0.95    &    0.267   &  0.38e-01  \\
J232659.08$-$002352.4   &  2.1598    &  3.98  &    0.24$^{a}$  &   8.36   &    45.52   &   -0.96  &   0.24   &  0.98,1.29    &    0.142   &  0.78      \\
J232734.73$+$002234.0   &  1.4939    &  2.56  &    0.06        &   9.27   &    46.20   &   -1.19  &   0.48   &  0.23,0.86    &    0.501   &  0.71e-05  \\
J234023.66$-$005326.9   &  2.0853    &  2.38  &   -0.30$^{a}$  &   9.10   &    46.96   &   -0.25  &   0.28   &  0.54,1.34    &    0.536   &  0.70e-05  \\
J235409.17$-$001947.9   &  0.4615    &  3.13  &   -0.03$^{a}$  &   9.28   &    45.44   &   -1.96  &   0.73   &  0.38,1.00    &    0.321   &  0.10e-01  \\
J235936.81$-$003112.7   &  1.0956    &  4.01  &   -0.51$^{a}$  &   8.12   &    45.34   &   -0.89  &   1.19   &  0.64,1.05    &    0.571   &  0.13      \\
\hline
\end{tabular}
\end{table}

\end{document}